# A gate defined quantum dot on the two-dimensional transition metal dichalcogenide semiconductor WSe$_2$

Xiang-Xiang Song,[1,2] Di Liu,[1,2] Vahid Mosallanejad,[1,2] Jie You,[1,2] Tian-Yi Han,[1,2] Dian-Teng Chen,[1] Hai-Ou Li,[1,2] Gang Cao,[1,2] Ming Xiao,[1,2] Guang-Can Guo,[1,2] and Guo-Ping Guo[1,2]

[1] *Key Laboratory of Quantum Information, CAS, University of Science and Technology of China, Hefei, Anhui 230026, China*
[2] *Synergetic Innovation Center of Quantum Information & Quantum Physics, University of Science and Technology of China, Hefei, Anhui 230026, China*

ABSTRACT

Two-dimensional layered materials, such as transition metal dichalcogenides (TMDCs), are promising materials for future electronics owing to their unique electronic properties. With the presence of a band gap, atomically thin gate defined quantum dots (QDs) can be achieved on TMDCs. Here, standard semiconductor fabrication techniques are used to demonstrate quantum confined structures on WSe$_2$ with tunnel barriers defined by electric fields, thereby eliminating the edge states induced by etching steps, which commonly appear in gapless graphene QDs. Over 40 consecutive Coulomb diamonds with a charging energy of approximately 2 meV were observed, showing the formation of a QD, which is consistent with the simulations. The size of the QD could be tuned over a factor of 2 by changing the voltages applied to the top gates. These results shed light on obtaining smaller quantum dots on TMDCs with the same top gate geometry, compared to traditional GaAs/AlGaAs heterostructure, for further researches.

*Correspondence and requests for materials should be addressed to G.P.G. (gpguo@ustc.edu.cn)

# I. INTRODUCTION

Two-dimensional layered materials, such as graphene or transition metal dichalcogenides (TMDCs), are considered as promising materials for future electronics owing to their unique electronic properties[1-3]. Compared with traditional semiconducting materials, two-dimensional layered materials have the advantages of their inherent flexibility, atomically thin geometry and dangling-bond-free interfaces, making them easy to integrate with various substrates. Graphene has been studied in high-speed electronic applications because of its ultrahigh mobility[4]. A variety of electronic applications have been demonstrated using graphene[5-7].

However, the absence of a band gap limits the performance of graphene-based devices, typically insufficient current on-off ratios[8]. Consequently, significant effort has been devoted to identifying alternative two-dimensional semiconductors. TMDCs are one of the most promising materials among those studied as they have band gaps ranging from 1 to 2 eV[9, 10]. A variety of devices made out of TMDCs have been demonstrated, including field-effect transistors (FETs)[11-17], heterostructure junctions[18, 19] and photodetectors[20, 21].

Meanwhile, from the quantum device aspect, especially for quantum dots (QDs), monolayer graphene cannot be electrically confined to form QDs because of the lack of a band gap. Plasma etching is widely used to shape graphene flakes to obtain QDs, both single dots[22, 23] and double dots[24-26]. However, the edge states and charge inhomogeneities induced by etching change the properties of the graphene flakes, limiting the performance of the nano-devices[27, 28], especially the coherency of the graphene QDs. With a band gap, TMDCs overcome the disadvantages associated with forming graphene QDs. Similar to traditional semiconductor heterostructures, QDs can be formed on atomically thin two-dimensional layered TMDC materials by applying a gate voltage to tune the local band structure. Recently, due to strong spin-orbital coupling, theoretical predictions using TMDC QDs as qubits have been proposed[29].

Semiconducting TMDCs consist of hexagonal layers of metal atoms (M) sandwiched between two layers of chalcogen atoms (X), which can be written as $MX_2$, where M = Mo or W and X = S, Se or Te[30, 31]. As one of the TMDCs, $WSe_2$ has a direct band gap of 1.7 eV for a monolayer and an indirect band gap of 1.2 eV for a bulk crystal[9]. Several studies have been performed, showing that the typical carrier mobility of $WSe_2$ is the range ~140–500 $cm^2/(V\ s)$[11, 14, 32].

Here, we used standard semiconductor fabrication techniques to demonstrate a quantum confined structure made of $WSe_2$, one of the TMDCs, with tunnel barriers defined by electric fields. Different from the common graphene etched QDs and graphene-like etched $MoS_2$ nanoribbons[33], the $WSe_2$ QDs were formed by applying a voltage to the top gates to decrease the influence of the edge states. Over 40 consecutive Coulomb diamonds were observed, demonstrating the formation of QDs that was predicted in our simulations. Furthermore, it is demonstrated that the size of the QDs could be tuned over a factor of 2 by changing the voltage applied to the top gates. Compared to traditional GaAs/AlGaAs heterostructure, for the same top gate geometry and voltages, the smaller quantum dot can be formed in layered material structure. These gate defined QDs in two-dimensional TMDCs open an avenue to explore the electronic properties of TMDCs. Also, tunable TMDC QDs that are isolated from the edge states could be used as spin or valley qubits in electron manipulations in further research studies.

## II. DEVICE FABRICATION

Similar to the graphene devices, the $WSe_2$ flakes were produced by mechanically cleaving a bulk $WSe_2$ crystal and depositing it onto a highly doped silicon substrate covered by 100 nm of $SiO_2$. The silicon substrate was used as the back gate. Few-layer flakes were selected using an optical microscope.

After depositing the $WSe_2$ flake on the substrate, we located where the $WSe_2$ flake precisely was and fabricated metal marks for further steps. The standard electron beam lithography (EBL) process was used to form the pattern of source-drain electrodes. We used E-beam evaporation of 10 nm Pd and 90 nm Au to fabricate the source-drain contacts. Then, an atomic layer deposition (ALD) technique was used to make a 40 nm thick insulating $Al_2O_3$ layer. Another EBL step followed by E-beam evaporation was used to fabricate four split top gates. 5 nm Ti and 45 nm Au was deposited to form the top gates. After the $Al_2O_3$ layer covering the source-drain contacts was etched, the device was bonded to the chip carrier for the measurements.

Figure 1(a) shows a scanning electron microscope (SEM) image of the $WSe_2$ device used in our experiments. The $WSe_2$ flake had an area of approximately 2 μm×3 μm. The thickness of the flake was measured using an atomic force microscope (AFM). Figure 1(b) shows that the thickness of the $WSe_2$ flake was 4.5 nm, corresponding to 7 layers[34]. Figure 1(c) shows a schematic cross-section of the device. The heavily doped silicon substrate operated as the back gate, which was isolated from the $WSe_2$ by a 100 nm thick layer of $SiO_2$. Depositing metal contacts on TMDCs are a critical issue for the electronic applications of TMDCs. Several studies have focused on reducing the Schottky barrier to form Ohmic contacts between the metal contacts and TMDCs[32, 35-37]. Here, Pd was used as the interlayer between Au and $WSe_2$ to obtain Ohmic contacts, these are labeled as "source" and "drain" in Figure 1(c). On the $Al_2O_3$ insulating layer, four top gates were fabricated to form the QD, these are labeled MG, LB, PG and RB. To obtain a better understanding of the geometric design, we used a commercial finite element analysis simulation tool (COMSOL) to calculate the potential profile in the $WSe_2$ layer in the top gate geometry. The potential profile of the structure used in the experiments (see Figure 1(a) and (c)) was modelled by solving the Poisson equation. As shown in Figure 1(d), the voltage applied to the top gates changed the local potential and confined the electrons. The closed contours indicate where the QD could exist. The experiments were performed in a He3 refrigerator at a base temperature of 240 mK. The standard Lock-in method was used to probe the electronic signals.

## III. RESULTS AND DISCUSSION

First, a DC voltage was applied to the back gate to tune the Fermi energy of the $WSe_2$ device. As the back gate voltage ($V_{BG}$) was swept the source-drain current was measured. Figure 2(a) shows that a non-zero current appeared at a back gate voltage of 36 V and increased when the back gate voltage was tuned to be more positive, showing the characteristic behavior of an n-doped semiconductor. Because an impurity potential could exist, which may modulate the local conduction band gap, affecting the QDs, the applied back gate voltage was tuned to more positive to make the device operate at a higher Fermi level in order to decrease the influence of the disorder potential[33]. We measured the current as a function of the DC bias voltage at different back gate voltages. As shown in Figure 2(b), changing the back gate voltage from 41.5 V to 45.5 V, over 40 consecutive Coulomb diamonds were observed, showing a QD formation. Here, all of the top gates had an applied DC voltage of −2 V to tune the local Fermi level underneath the electrodes.

By zooming in the region highlighted with red dotted lines in Figure 2(b), the $E_C$ of the $WSe_2$ QDs could be measured. The charging energy $E_C$ remained almost constant from $V_{BG}$=43.1 V to $V_{BG}$=44.1 V, as shown in Figure 2(c). The $E_C$ was estimated to be approximately 2 meV. The lever arm of the back gate ($\alpha_{BG}$) was 0.020 eV/V. Figure 2(d) shows the current as a function of back gate voltage, obtained along the red dashed line in Figure 2(c), known as Coulomb peaks.

Next, to investigate whether the QD was formed by the electric field applied to the top gates, the voltage applied to the top gates was changed to tune the size of the $WSe_2$ QD. For simplicity, the same voltage was applied to all of the top gates. We measured the Coulomb diamonds as a function of the back gate voltage, as described above, at different top gate voltages. The $E_C$ for over 20 Coulomb diamonds was measured from $V_{BG}$=42 to 45 V for each top gate voltage. Statistic distributions of the $E_C$ are shown in Figure 3(a)–(f). Here, the top gate voltage was tuned from −4 V to 1 V. We found that when the top gate voltage ($V_{TG}$) became more negative, larger $E_C$ appeared, indicating that the QDs had become smaller. When the $V_{TG}$ was changed more negative to −2 V, larger $E_C$ (above 2.5 meV) was observed, suggesting the size of the QD decreased. The number of Coulomb diamonds, which had an $E_C$ larger than 2.5 meV, increased if we continued to tune $V_{TG}$ to more negative. The average $E_C$ was measured as a function of $V_{TG}$, shown as the black squares in Figure 3(g). $E_C$ increased when $V_{TG}$ was changed from 1 V to −4 V. Using $E_C=e^2/(8\varepsilon_0\varepsilon_r r)$ [38], where $\varepsilon_r$ is the relative permittivity of $WSe_2$ and r is the radius of the QD, it was estimated that the QD radius changed from 275 nm to 150 nm, as shown with the blue circles in Figure 3(g). The area of each QD (~$r^2$) decreased by over a factor of 2 when $V_{TG}$ became more negative, suggesting that the top gates tuned the confinement potential that formed the QD. Note that, because the estimated dot size is comparable to the thickness of the insulating $SiO_2$ layer, the isolated disk approximation should change to a parallel plate approximation. Thus, the result of the estimated QD size should be considered as an upper limit, which is comparable to our geometric design.

We also investigated the change in the dot size qualitatively, using COMSOL simulations. Figure 3(h)–(j) show the potential profiles of the areas where closed the contours existed in Figure 1(d). By changing the voltages applied to the top gates from 0 V to −4 V, we plotted the potential profiles at different $V_{TG}$ in the same range. The confinement potential became lower and the dot size decreased when $V_{TG}$ was tuned to be more negative, which is consistent with the experimental results. In addition, there were some variations in the size of Coulomb diamonds, as shown in Figure 3(a)–(f). As the $V_{BG}$ was tuned monotonically, which can be understood as changing the Fermi level, the $E_C$ should also change monotonically if the energy band varies smoothly. The variations in the $E_C$ suggest that fluctuations may exist in the energy band which may be caused by local impurities. Such local impurities modified the band structure of the flake, resulting in variations in $E_C$. Similar results were observed in Ref. [33]. The influence of the local impurity potential had not been excluded, although the devices worked at a relatively high $V_{BG}$.

Furthermore, to demonstrate the direct influence of the top gate, the plunger gate was tuned to obtain Coulomb diamonds. It is shown in Figure 4 that when the plunger gate voltage was changed from −1.7 V to −2.2 V, Coulomb diamonds were observed. Note that the plunger gate still had an influence on the confinement potential, as $E_C$ became larger when the voltage of the plunger gate was tuned to be more negative. This result is consistent with the result shown in Figure 3(g). We estimated the lever arm of the plunger gate ($\alpha_{PG}$) to be about 0.048 eV/V here. The difference between $\alpha_{BG}$ and $\alpha_{PG}$ can be calculated using: $\frac{\alpha_{PG}}{\alpha_{BG}} = \frac{\varepsilon_{PG}}{\varepsilon_{BG}} \times \frac{d_{BG}}{d_{PG}}$, where $\varepsilon_{BG(PG)}$ is the relative dielectric constant of the back gate (plunger gate) and $d_{BG(PG)}$ is the distance between the dot and the back gate (plunger gate). Using the $\alpha_{BG}$ value obtained above, the calculated $\alpha_{PG}$ should be 0.094 eV/V, which is about 2 times larger than the $\alpha_{PG}$ measured here. This can be understood because the relative dielectric constant of $Al_2O_3$ may be smaller, caused by the quality of the growth.

Here, we found some evidence of overlapped Coulomb diamonds which suggests double dots may be formed, shown by the black arrows in Figure 4. Since the edge states were absent in our devices, we believed that there are two possibilities that may cause the formation of the double dots. One is the aggressive gate voltage applied to the top gates, which changed the electron distribution. The other is disorders in the crystals, which trap impurities, affecting the confinement potential. These disorders may also contribute to another conductance channel, which results in the non-zero Coulomb blockade current observed in Figure 2(d). However, different from the edge states, this impurity problem could be solved in principle, when the crystal-growth technique improves.

Next, the source and drain contacts in the devices will be discussed. Figure 5(a) shows a schematic of two Coulomb diamonds. Using the constant interaction model, the slope of the two sides can be calculated as $|e|C_g/C_S$ and $-|e|C_g/(C-C_S)$ respectively[39], as shown in Figure 5(a). Here, $C_{S(D)}$ is the capacitance between the source (drain) and the dot, $C_g$ is the capacitance between the gate and the dot, and C is the total capacitance: $C=C_S+C_g+C_D$. In our experiments, the Coulomb diamonds were always asymmetric. Moreover, when the top gates voltages were tuned to a different region, the Coulomb diamonds exhibited different configurations, as shown in Figure 5(b) and (c). In Figure 5(b), the Coulomb diamonds have one side almost parallel to the y-axis, indicating that $C_S$ was almost zero. However, in Figure 5(c), the side of the Coulomb diamonds which parallels to the y-axis changed to the other side, suggesting

$C_S$ became larger as the domination of the total capacitance C in this configuration ($C=C_S$). The change in $C_S$ suggests that the location of the WSe$_2$ dot varied under the gate potential. However, due to the limitation of the accuracy of estimation from diagram, it is hard to get quantitative estimation such as the exact location of the dot. Note that $C_S$ is much larger than $C_g$ in the situation in Figure 5(c), as $C=C_S$. We can estimate the source and drain contacts of our device were not purely Ohmic. Although Schottky barriers existed on the device, the quantum confinement is still valid. In order to achieve further manipulation of the electrons in the QDs on TMDCs, more studies are needed to improve the contact between the TMDCs and metal electrodes.

Finally, we want to discuss the difference between our layered material gate defined quantum dot and traditional GaAs/AlGaAs heterostructure quantum dot. Because of the existence of a layer of donors, the back gate voltage $V_{BG}$ is no longer needed here to accumulate carriers. Using COMSOL, we can simulate the potential profile at the interface of AlGaAs and GaAs, where 2DEG (2 dimensional electron gas) is formed. As shown in Figure 6, the potential profile for GaAs/AlGaAs heterostructure with the same top gate geometry and $V_{TG}$ is calculated. Compared Figure 6 to Figure 1(d), the closed contours disappear, which indicates the quantum dot is no longer formed in this case. Moreover, the potential in the area between barriers (LB, RB) and plunger gate (PG) is in the same range as in the area surrounded by the top gates. This result is different from Figure 1(d), where the potential is pulled up in the area between barriers (LB, RB) and plunger gate (PG), indicating that the electrons are depleted underneath the gates (LB, RB, PG). While in the case for Figure 6, electrons still can tunnel through the channel between the gates (LB, RB, PG). This means the distance between the top gates needs to be decreased in order to confine a quantum dot when applying the same $V_{TG}$ for the GaAs/AlGaAs heterostructure, which increases the difficulty in fabrication steps. Alternatively, for the same top gate geometry and $V_{TG}$, the smaller quantum dot with lager energy of excited states could be formed in layered material structure, which is essential for further studies. Moreover, changing the Al$_2$O$_3$ with another layered material h-BN, we can decrease the thickness of insulator layer further to even one layer of atoms, allowing a more aggressive influence of top gates. While for GaAs/AlGaAs heterostructure, obtaining 2DEG forbids decreasing the vertical distance between the top gates and electrons. This difference allows fabricating atomic-scale and even flexible quantum dots to have a better control of electrons inside.

## IV. CONCLUSIONS

In summary, standard semiconductor fabrication techniques were used to obtain a gate defined WSe$_2$ QD with tunnel barriers defined by electric fields. Over 40 consecutive Coulomb diamonds with an $E_C$ of approximately 2 meV were observed, showing the confinement potential that was predicted by our simulations. By tuning the gate voltage applied to both of the top gates and the back gate, we observed the statistic distribution of the $E_C$ varied, indicating that the size of the dots changed by over a factor of 2. For the same top gate geometry and $V_{TG}$, the smaller quantum dot can be formed in layered material structure, compared to traditional GaAs/AlGaAs heterostructure, where quantum dots may even vanish for our top gates design. These gate defined QDs in two-dimensional TMDCs open an avenue to explore the electronic properties of TMDCs. Also, tunable TMDC QDs, which are isolated from the edge states, shed light on quantum nano-devices in TMDCs for further research studies.

## ACKNOWLEDGEMENTS

This work was supported by the National Fundamental Research Program (Grant No. 2011CBA00200), the National Natural Science Foundation (Grand Nos. 11222438, 11174267, 61306150, 11304301 and 91121014) and the Chinese Academy of Sciences.

Figures:

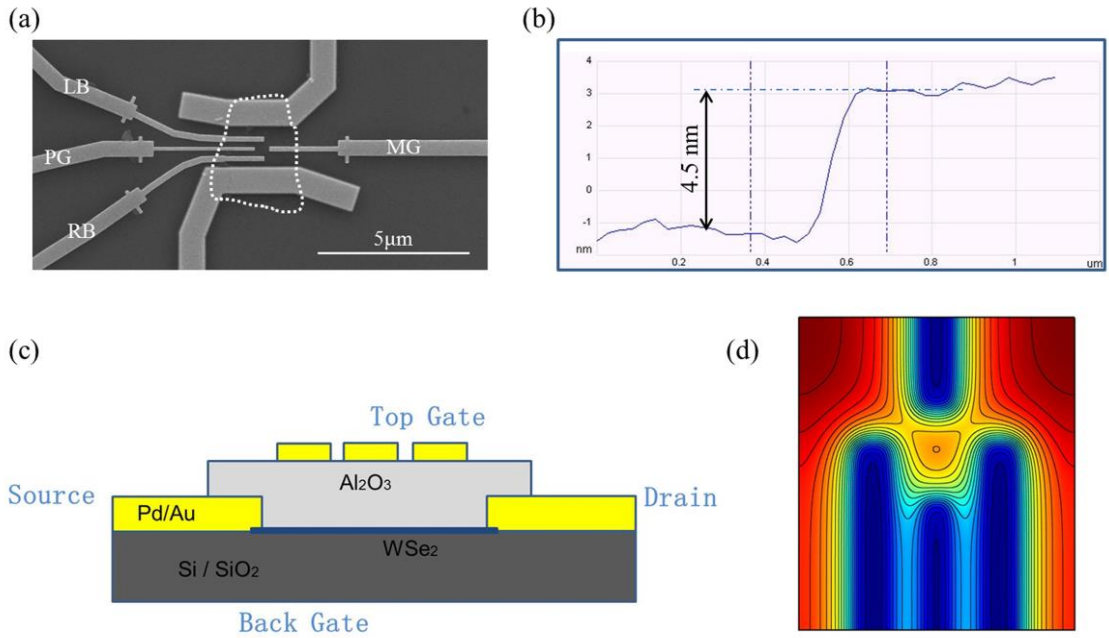

FIG. 1. (a) SEM image of the WSe$_2$ QD studied in this work. The WSe$_2$ flake is highlighted by the white dotted line. The four top gates are labeled as MG, LB, PG and RB. The white bar has a length of 5 μm. (b) Thickness profile of the WSe$_2$ flake (4.5-nm thick) measured with an AFM. (c) Schematic cross-section of the device. The few-layer WSe$_2$ was deposited on a heavily doped silicon substrate covered with 100-nm-thick SiO$_2$. The WSe$_2$ flake was separated from the four top gates (Ti/Au) by 40 nm of ALD-grown Al$_2$O$_3$. Two metal gates (Pd/Au) were connected to the flake used as the source-drain contacts. (d) COMSOL simulations on the potential profile in the WSe$_2$ layer for the gate pattern of the device shown in (a). Here $V_{MG}=V_{LB}=V_{PG}=V_{RB}=-2$ V and $V_{BG}=43$ V. The closed contours indicate where the QD could exist.

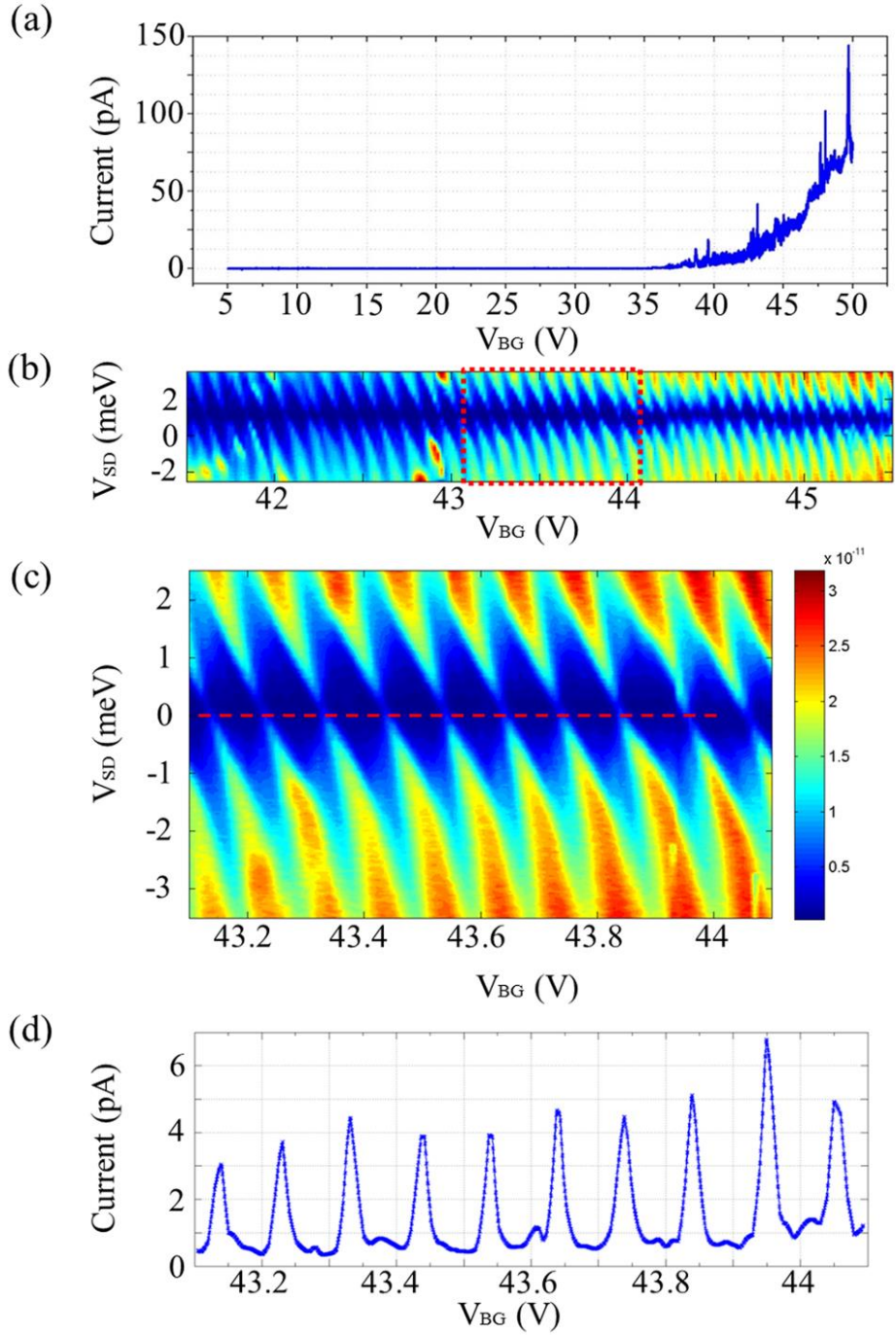

FIG. 2. (a) Source-drain current flow through the WSe$_2$ device as a function of back gate voltage (V$_{BG}$), showing the characteristic behavior of an n-doped semiconductor. (b) Over 40 consecutive Coulomb diamonds were measured from V$_{BG}$=41.5 V to 45.5 V. All of the top gates had an applied DC voltage of −2 V. (c) Zoom in of the region highlighted by the red dotted lines in (b), showing that the E$_C$ was approximately 2 meV. (d) A series of Coulomb peaks measured along the red dashed line in (c).

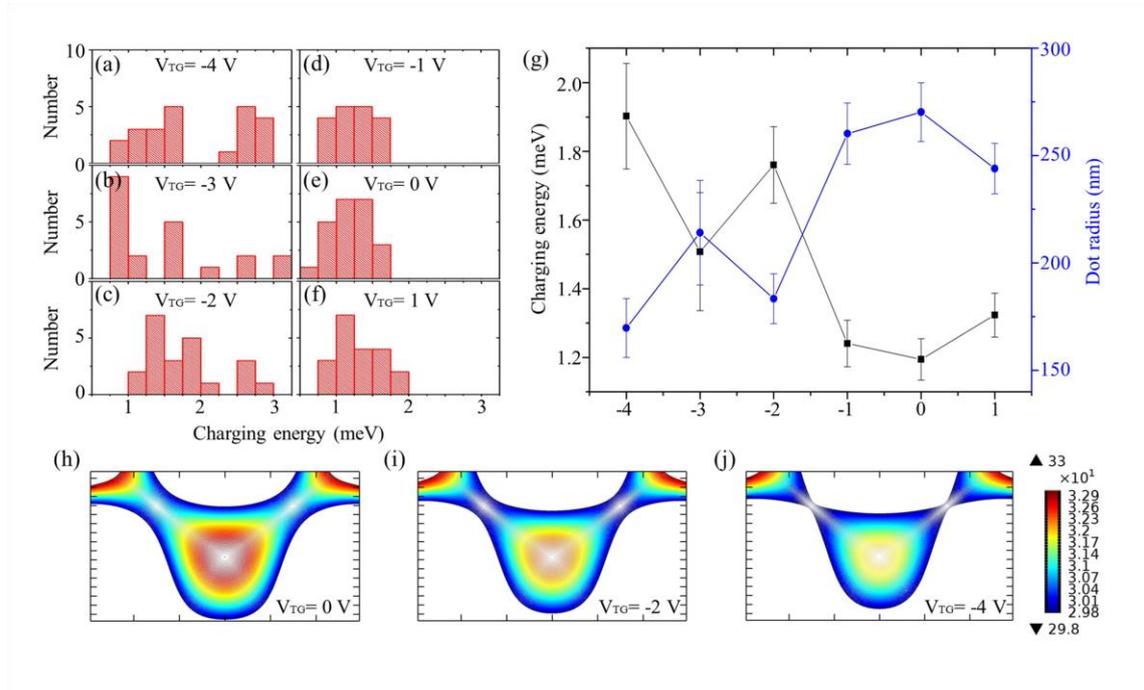

FIG. 3. (a)–(f) Statistic distributions of the $E_C$, measured from over 20 Coulomb diamonds in the range from $V_{BG}$=42 V to 45 V at different top gate voltages ($V_{TG}$s). (g) Average $E_C$ (black squares) and calculated quantum radii (blue circles) as a function of $V_{TG}$. (h)–(j) Potential profile of the area where the closed contours were in Figure 1(d) at different $V_{TG}$ values, showing the change in the dot size. Here $V_{BG}$=43 V.

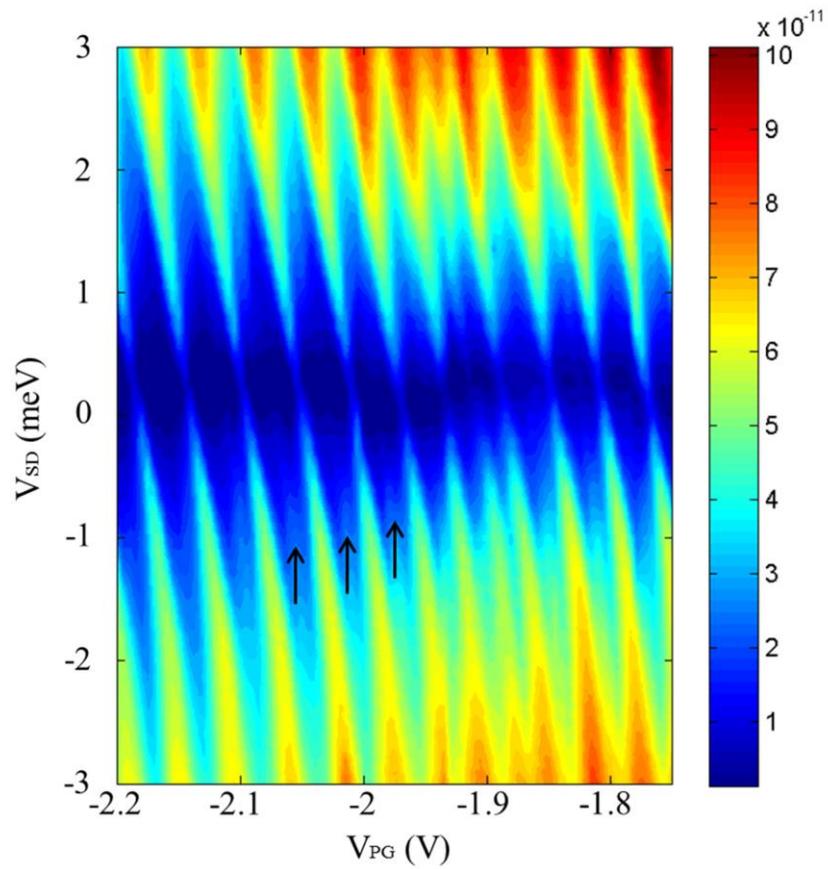

FIG. 4. Coulomb diamonds measured as a function of the gate voltage applied to the plunger gate, from $V_{PG}=-2.2$ V to $-1.7$ V. The $E_C$ changed with $V_{PG}$. Here $V_{MG}=V_{LB}=V_{RB}=-2$ V and $V_{BG}=44.5$ V. The black arrows indicate evidence that double dots may have formed.

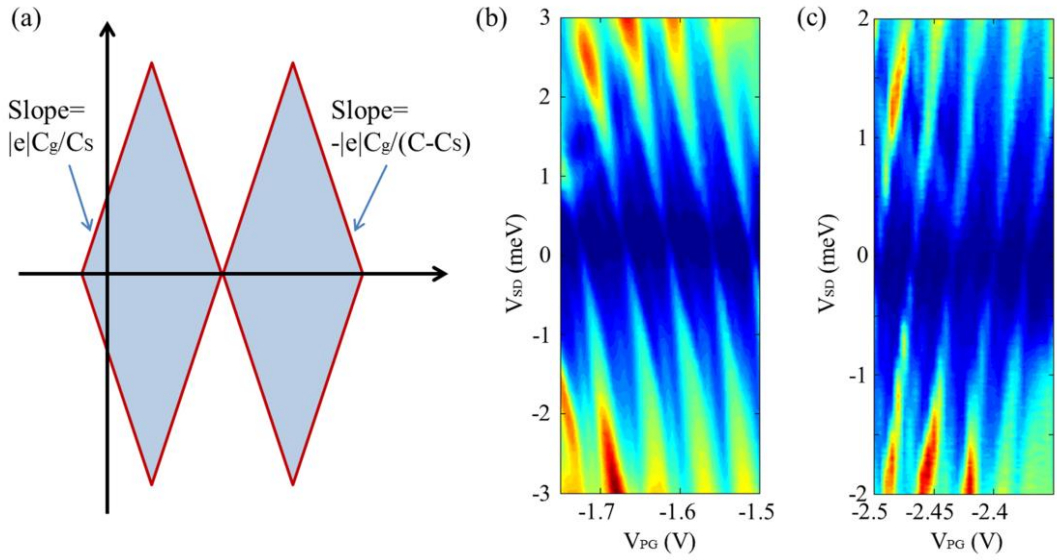

FIG. 5. (a) Schematic of two consecutive Coulomb diamonds. The slopes of the two sides are $|e|C_g/C_S$ and $-|e|C_g/(C-C_S)$. (b) Coulomb diamonds obtained from $V_{PG}=-1.75$ V to $-1.5$ V. Here, $V_{RB}=-2$ V, $V_{MG}=V_{LB}=-1.5$ V and $V_{BG}=42.5$ V. (c) Different configurations of the Coulomb diamonds obtained from $V_{PG}=-2.5$ V to $-2.35$ V. Here $V_{RB}=-2.15$ V, $V_{MG}=V_{LB}=-2.25$ V and $V_{BG}=43$ V.

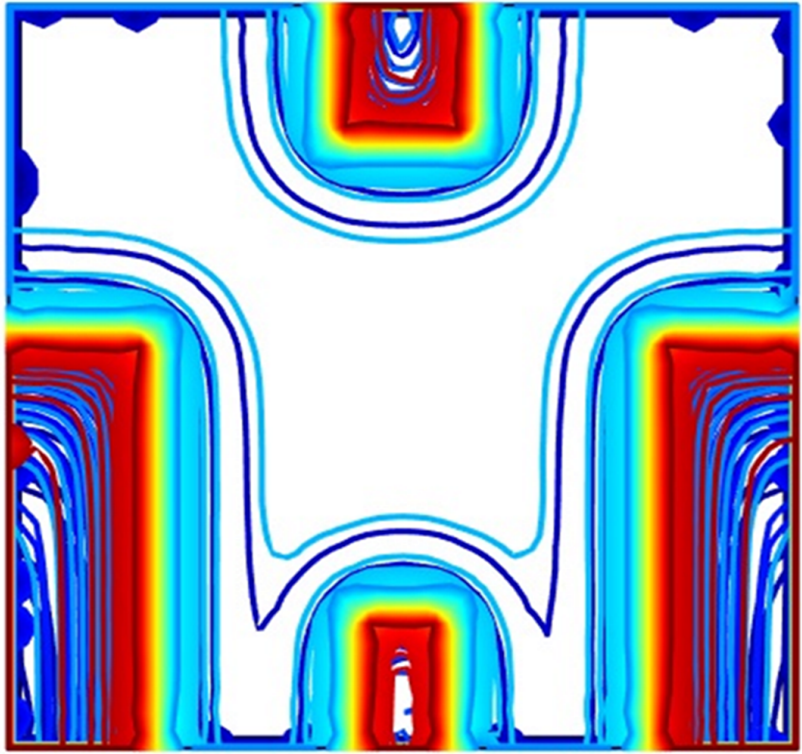

FIG. 6. COMSOL simulations on the potential profile for GaAs/AlGaAs heterostructure, with the same $V_{TG}$ and top gate geometry of the device shown in Figure 1(d), is calculated. Here $V_{MG}=V_{LB}=V_{PG}=V_{RB}=-2$ V. The closed contours shown in Figure 1(d) disappear, indicating the QD no longer exists in this case. In the simulation, the distance between the top gates and 2DEG is about 95 nm, including a 10 nm of GaAs cap and an 85 nm of AlGaAs.